\documentclass{rmaa}
\nonstopmode

\usepackage[latin1]{inputenc}

\title{A Lossy Method for Compressing Raw CCD Images}

\author{Alan M. Watson
\affil{Instituto de Astronom{\'\i}a,
Universidad Nacional Aut\'onoma de M\'exico,
Campus Morelia, México}}

\fulladdresses{
\item Alan. M. Watson: Instituto de Astronom{\'\i}a, 
Universidad Nacional Aut\'onoma de M\'exico,
Campus Morelia, 
Apartado Postal 3-72 (Xangari)
58089 Morelia, Michoac\'an, M\'exico. (a.watson@astrosmo.unam.mx)
}

\shortauthor{Watson} 
\shorttitle{Compressing Raw Images}

\resumen{Se presenta un método para comprimir las imágenes
crudas de dispositivos como los CCD. El método es muy
sencillo: cuantización con pérdida y luego compresión sin
pérdida con herramientas de uso general como gzip o bzip2.
Se converten los archivos comprimidos a archivos de FITS
descomprimiéndolos con gunzip o bunzip2, lo cual es una
ventaja importante en distribuir datos comprimidos. El grado
de cuantización se elige para eliminar los bits de bajo
orden los cuales sobre muestrean el ruido, no proporcionan
información, y son difíciles o imposibles comprimir. El
método es con pérdida, pero proporciona ciertas garantías
sobre la diferencia absoluta máxima, la diferencia RMS, y la
diferencia promedio entre la imagen comprimida y la imagen
original; tales garantías implica que el método es adecuado
para comprimir imágenes crudas. El método produce imágenes
comprimidas de 1/5 del tamaño de las imágenes originales
cuando se cuantizan imágenes en las que ningún valor cambia
por más de 1/2 de la desviación estándar del fondo. Esta es
una mejora importante con respecto a las razones de
compresión producidas por métodos sin pérdida, y parece que
las imágenes comprimidas con bzip2 no exceden el límite
teórico por más de unas decenas de porciento. }

\abstract{ This paper describes a lossy method for
compressing raw images produced by CCDs or similar devices.
The method is very simple: lossy quantization followed by
lossless compression using general-purpose compression tools
such as gzip and bzip2. A key feature of the method is that
compressed images can be converted to FITS files simply by
decompressing with gunzip or bunzip2, and this is a
significant advantage for distributing compressed files. The
degree of quantization is chosen to eliminate low-order bits
that over-sample the noise, contain no information, and are
difficult or impossible to compress. The method is lossy but
gives guarantees on the maximum absolute difference, the
expected mean difference, and the expected RMS difference
between the compressed and original images; these guarantees
make it suitable for use on raw images. The method
consistently compresses images to roughly 1/5 of their
original size with a quantization such that no value changes
by more than 1/2 of a standard deviation in the background.
This is a dramatic improvement on lossless compression. It
appears that bzip2 compresses the quantized images to within
a few tens of percent of the theoretical limit. }


\begin{document}

\maketitle

\section{Introduction}

Optical and infrared instruments now routinely produce huge
amounts of image data. The largest current common-user CCD
mosaic has $\rm 12k \times 8k$ pixels (Veillet 1998); a
single image from such a mosaic is 192 MB in size. The
largest current common-user infrared detector mosaic has
$\rm 2k \times 2k$ pixels (Beckett et al.\ 1998), but makes
up for its smaller size by being read more frequently. More
than a few nights of data from such instruments can easily
overwhelm workstation-class computers. Compression is a
solution to some of the problems generated by these large
quantities of data. Similarly, compression can improve the
effective bandwidth to remote observatories (in particular
space observatories), remote data archives, and even local
storage devices.

This paper describes in detail a lossy method for
compressing raw images and presents a quantitative
comparison to other lossy methods. It is organized as
follows: \S~2 reviews the limitation of lossless compression
and the motivation for lossy compression; \S~3 briefly
summarizes the most relevant previous work on lossy
compression; \S~4 describes the new method; \S~5 presents
results on the distribution of differences between the
original and compressed images; \S~6 investigates the
performance of the method with particular reference to
hcomp; \S~7 compares the performance of the method to other
similar methods; \S~8 discusses the suitability of the
method for compressing raw data and investigates some
consequences of such use; \S~9 discusses how the method
might be improved; and \S~10 presents a brief summary.

\section{Lossless Compression}

\label{section-pure-gaussian-noise}

\subsection{The Shannon Limit}

Compression methods can be classified as lossless or lossy
depending on whether compression changes the values of the
pixels. The difficulties of lossless compression of CCD
images have been discussed by White (1992), Press (1992),
and V\'eran \& Wright (1994). The noise in images from an
ideal CCD consists of Poisson noise from the signal and
uncorrelated Gaussian read noise. If the signal is measured
in electrons, the variance of the Poisson noise is equal to
the signal and the variance of the read noise is typically
10--100. However, CCD images are often quantized to sample
the read noise, with typical analog-to-digital converter
gains being 1--5 electrons. This means that the larger
noises associated with larger signal levels can be hugely
over-sampled. If this is the case, the low-order bits in an
image are essentially uniform white noise, contain no
information, and cannot be compressed.

To illustrate this, series of $512 \times 512 \times
\mbox{16-bit}$ FITS images of pure Gaussian noise were
created. The images were written with with a BSCALE of
$q\sigma$, so the standard deviation is sampled by a factor
of $1/q$. Shannon's first theorem (Shannon 1948ab, 1949)
allows us to calculate the optimal compression ratio for
these images. The theorem states that if a stream of bits is
divided into fixed-length ``input code units'', then the
minimum number of bits required to encode each input code
unit is just the Shannon entropy,
\begin{equation}
H \equiv -\sum_{\forall p_i \ne 0} p_i \log_2 p_i,
\label{equation-shannon-entropy}
\end{equation}
where $p_i$ is the normalized frequency of the input code
unit $i$. From this we can derive that the optimal input
code unit is the longest one for which the bits are still
correlated; since in this case, each pixel is independent,
the optimal input code unit is simply a 16-bit pixel. The
resulting optimal compression ratio (the ratio of the
compressed size to original size) will be $H/16$.

\begin{figure}[t]
\begin{center}
\includegraphics[height=\linewidth,angle=270]{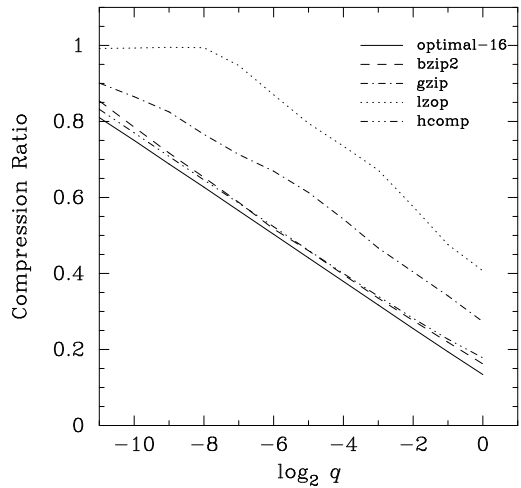}
\end{center}
\caption{The compression ratio (ratio of compressed size to
original size) as a function of $\log_2 q$ for an image
consisting of pure Gaussian noise with standard deviation
$\sigma$ written as a 16-bit FITS image with a BSCALE of
$q\sigma$ and then losslessly compressed. The solid line is
the Shannon limit for optimal compression of the individual
16-bit pixels and the other lines show the ratios achieved
by bzip2, gzip, lzop, and hcomp (used losslessly). }
\label{figure-random-bits}
\end{figure}

Figure \ref{figure-random-bits} shows as a function of $q$
the optimal compression ratio determined by calculating the
Shannon entropy explicitly. As can be seen, the optimal
compression ratio is a constant minus $(\log_2 q)/16$.
Intuitively, this is expected: the number of incompressible
low-order bits grows as $-\log_2 q$ and the fraction of the
compressed image occupied by these bits grows as $-(\log_2
q)/16$. This result is not new; Romeo et al.\ (1999)
investigated the Shannon limit for compression of a
quantized Gaussian distribution, and their equation (3.3),
in my notation, is
\begin{equation}
H \approx \log_2\sqrt{2\pi e} - \log_2 q,
\label{equation-shannon-entropy-for-gaussian}
\end{equation}
This equation was derived in the limit of small $q$, but
Gaztañaga et al.\ (2001) have shown that the corrections for
finite $q$ are small for $q$ at least as large as 1.5. Not
surprisingly, the Shannon entropies and compression ratios
calculated using this equation and calculated explicitly for
the Gaussian noise images are in almost perfect agreement.

Real astronomical images contain information in addition to
noise, but nevertheless this exercise demonstrates the
futility of attempting to achieve good compression ratios
for images that contain over-sampled noise. For example, if
the noise is over-sampled by a factor of 60 (which
corresponds to $q = 60$), it is \emph{impossible} to
compress the image losslessly to better than half its size;
too much of the image is occupied by white noise in the
low-order bits. This exercise also explains why lossless
methods are often more successful on images with low
signals, such as biases or short exposures, and less
successful on images with high signals, such as flats, deep
exposures, and infrared images. 

We are forced to the conclusion that if we wish to
significantly and consistently compress raw images,
including those with high background levels, there is no
alternative but to use lossy compression
methods.\footnote{The results of White \& Becker (1998)
appear to contradict this statement. These authors achieve
compression ratios of 17\%--32\% on four WFPC2 images using
lossless Rice compression. However, WFPC2 images have low
backgrounds, due to the small pixels of WFPC2 and low sky
brightness in orbit, and the gain normally under-samples the
read noise. It is likely that none of their images has a
large white noise component, and as such their sample of
images is not representative of the full range of raw
astronomical images.}

\subsection{Real Lossless Compression Methods}

The Shannon limit is a theoretical limit and is independent
of any particular real compression method. Nevertheless, it
is useful to consider how well real lossless compression
methods perform on Gaussian noise both to introduce these
methods and to investigate how well they perform in this
idealized case. The methods considered are Huffman coding,
Arithmetic coding, hcomp, bzip2, gzip, and lzop.

Huffman coding (Huffman 1952; Press et al.\ 1992, \S~20.4)
and Arithmetic coding (Whitten, Neal, \& Cleary 1987; Press
et al.\ 1992, \S~20.5) make use of the frequency
distribution of fixed input code units. As such, if 16-bit
pixel values are taken as the input code unit, the
compression ratio should be close to the Shannon limit
(provided the image is large compared to the frequency
table). Indeed, Gaztañaga et al.\ (2001) show that Huffman
coding of quantized Gaussian noise is very nearly optimal
for $q \le 1$ (see their Figure 1). For this reason, we need
not investigate actual implementations of these methods but
instead can approximate their compression ratios by the
Shannon limit.

Hcomp (White 1992) is specialized to compressing 16-bit
astronomical images and uses a complex algorithm based on a
wavelet transformation of the pixel values, optional
quantization of the coefficients, and quad-tree compression
of the coefficients. Hcomp can be used losslessly by
omitting the quantization of the coefficients.

Bzip2 (Seward 1998), gzip (Gaily 1993), and lzop (Oberhumer
1998) are methods for compressing general byte streams and
use dictionary-based algorithms with variable-length input
code units. They are portable, efficient, and free from
patents. Sources and executables are freely available. They
represent different trade-offs between compression ratio and
speed, with bzip2 typically being slowest but achieving the
best compression ratios, lzop typically being the fastest
but achieving the worst compression ratios, and gzip being
intermediate in both speed and compression ratio.

Figure 1 shows the compression ratio achieved by hcomp (used
losslessly), bzip2, gzip, and lzop. The compression ratios
have the same trend as the Shannon limit -- they also cannot
compress white noise in the low order bits -- but the
constant is larger. (Lzop is a slight exception in that it
is initially unable to compress the image but then follows
the expected trend.) Both bzip2 and hcomp are quite close to
the Shannon limit. The surprisingly good performance of
bzip2 suggests that it might be useful in compressing
images; a large portion of the rest of this paper will be
devoted to evaluating this suggestion.

\section{Prior Work on Lossy Compression}

Louys et al.\ (1999) have recently scrutinized lossy
compression methods, including hcomp (White 1992). Their
investigation concentrated on relatively high compression of
reduced images. Compressing raw images is somewhat
different, in that we wish to maintain all of the
information in the image and avoid introducing spurious
features. Even weak features can be important, as subsequent
co-addition, spatial filtering, or other processing may
increase their significance. Therefore, for raw images we
require methods that suppress only those parts of an image
that contains no information (the white noise in the
low-order bits) and preserve perfectly those parts that do
contain information (the mid-order and high-order bits).
None of the lossy methods considered by Louys et al.\ (1999)
can give this guarantee. For raw images, we must consider
other methods.

More promising is the work of White and Greenfield (1999),
Nieto-Santisteban et al.\ (1999), and Gaztañaga et al.\
(2001). These authors have all presented lossy compression
methods that work by discarding white noise and then using a
lossless compression method. White and Greenfield (1999)
discard white noise from generic floating-point images by
quantizing each row by a fraction of the estimated noise and
converting to integers, Nieto-Santisteban et al.\ (1999)
discard white noise from simulated NGST data by scaling and
taking the square root, and Gaztañaga et al.\ (2001) discard
white noise from simulated data from cosmic microwave
background experiments by quantizing by a fraction of the
standard deviation. White and Greenfield (1999) and
Nieto-Santisteban et al.\ (1999) use the Rice compression
algorithm (Rice, Yeh, \& Miller 1993) and Gaztañaga et al.\
(2001) use Huffmann coding (Huffman 1952; see Press et al.\
1992, \S~20.4).

Comparisons to the work of White and Greenfield (1999) and
Nieto-Santisteban et al.\ (1999) are made in
\S~\ref{section-other-quantization-methods}; the method
suggested by Gaztañaga et al.\ (2001) is very similar to the
one suggested here, but has a completely different context.

\section{Method}
\label{section-method}

The quantization compression method is so simple that it is
somewhat embarrassing to describe it as a ``method''. It
consists of lossy quantization (or resampling) in brightness
of the original image to produce a quantized image, followed
by lossless compression of the quantized image.
Decompression consists of reversing the lossless compression
to recover the quantized image. The resampling quantum will
normally be a specified fraction of the expected noise in
the background. The aim of quantization is to reduce the
over-sampling of the noise, reduce the number of bits of
white noise, and permit the lossless compression method to
perform better. The quantization method is intended to be
used on raw data from CCD-like devices.

\subsection{Quantization}
\label{section-method-quantization}

The first stage of the quantization method is to produce a
quantized image. For definiteness, the quantization is
described in terms of integer FITS images (Wells, Greisen,
\& Harten 1981). This is the most common format for archived
raw data. The extension to other image formats is obvious.

Each integer $n$ in the original FITS image is related to a
data value $b$ by
\begin{equation}
b = b_{\rm zero} + n b_{\rm scale},
\end{equation}
where $b_{\rm zero}$ and $b_{\rm scale}$ are taken from the
BZERO and BSCALE records in the FITS header and have
defaults of 0 and 1. The data quantum $Q_b$ is chose to be
an integral multiple $Q_n$ of $b_{\rm scale}$. The quantized
brightness $\hat b$ is thus
\begin{equation}
\hat b = b_{\rm zero} + Q_b [b/Q_b] b_{\rm scale}.
\end{equation}
Here, $[x]$ is the nearest integer to $x$ with half integers
rounded to the nearest even integer to help avoid biases.
The quantized image is written with the same BZERO and
BSCALE values as the original image. Thus, the quantized
integer $\hat n$ in the quantized FITS file is related to`
the quantized data value $\hat b$ by
\begin{equation}
\hat n = (\hat b - b_{\rm zero}) / b_{\rm scale}.
\end{equation}
It can be seen that this quantization can be carried out
entirely in integer arithmetic as
\begin{equation}
\hat n = Q_n [n/Q_n].
\end{equation}
Certain values in the original image are special and are
treated accordingly. Thus, ``blank'' pixels (those which
have the value specified by the BLANK record in the FITS
header) are not changed and any pixel whose quantized value
would be the blank value is not changed. The values
corresponding to the maximum and minimum integer values are
treated similarly, as these values are often used to flag
saturated data or implicitly blank pixels. (Some images have
ancillary information, such as the exposure time and
engineering information, encoded in ``data'' pixels; this
information should be transferred to the header prior to
compression.)

The quantum $Q_b$ should be determined according to the
noise model for the detector. For CCD-like devices, the
estimated standard deviation $\sigma_b$ of the noise is
given by
\begin{equation}
\label{equation-noise-model}
\sigma_b = \left\{
\begin{array}{ll}
((b - b_{\rm bias}) g + r^2)^{1/2}/g    &b > b_{\rm bias}\\
r/g                                     &b \le b_{\rm bias}
\end{array}
\right..
\end{equation}
Here $g$ is the gain in electrons, $r$ is the read noise in
electrons, and $b_{\rm bias}$ is the bias level. We first
determine the bias level $b_{\rm bias}$ and data background
level $b_{\rm data}$ in the image, perhaps from the mean,
median, or mode in certain regions, and hence the noise by
applying Equation (\ref{equation-noise-model}). We then
determine the quantum as
\begin{equation}
\label{equation-x-quantization}
Q_b = \left\{
\begin{array}{ll}
b_{\rm scale} \lfloor q \sigma_b / b_{\rm scale} \rfloor
&q \sigma_b > b_{\rm scale}\\
b_{\rm scale}
&q \sigma_b \le b_{\rm scale}\\
\end{array}
\right.,
\end{equation}
or
\begin{equation}
Q_n = \left\{
\begin{array}{ll}
\lfloor q \sigma_b / b_{\rm scale} \rfloor
&q \sigma_b > b_{\rm scale}\\
1
&q \sigma_b \le b_{\rm scale}\\
\end{array}
\right..
\end{equation}
Here $\lfloor x \rfloor$ is the largest integer no greater
than $x$. Thus, the parameter $q$ controls the coarseness of
the quantization by setting the size of the quantum $Q_b$ to
be the smallest integer multiple of $b_{\rm scale}$ no
larger than $q$ times the noise. Suitable values for $q$ are
0.5 to 2 (see \S~\ref{section-differences} and
\S~\ref{section-performance}).

There is no requirement that the quanta be equal over the
entire image. The data and overscan regions of an image
should probably be quantized with different quanta as they
likely have different characteristic noises. Furthermore, if
the background and hence the noise vary significantly within
the data region, it makes sense to subdivide the data region
and use different quanta within each subregion. However, it
is important the quanta are constant within regions and are
not determined on a pixel-to-pixel basis, as this can
introduce small but nevertheless real biases (see
\S~\ref{section-other-quantization-methods}).

\subsection{Compression}
\label{section-method-compression}

The quantized FITS file is no smaller than the original FITS
file. Therefore, the second stage of the quantization method
is to compress the quantized FITS file using a lossless
compression method. The methods investigated here are bzip2
(Seward 1998), gzip (Gaily 1993), lzop (Oberhumer 1998), and
hcomp (White 1992). Hypothetical optimal 16-bit and 32-bit
encoders (which will approximate Huffman or Arithmetic
encoders) are also considered.

\subsection{Decompression}
\label{section-method-decompression}

The quantized image is recovered as a standard FITS file by
reversing the lossless compression using bunzip2, gunzip,
lzop, or hdecomp.

\section{Distribution of Differences}
\label{section-differences}

One advantage of the simplicity of the quantization method
is that we can derive useful information on the distribution
of the differences between the quantized and original
images. These give the confidence required to apply the
quantization method to raw data. Note that the distribution
of differences is determined only by the quantization stage;
the compression stage is lossless.

We define the normalized difference $\Delta$ between the
quantized and original data values to be
\begin{equation}
\label{equation:delta}
\Delta = {\hat b - b \over \sigma_b}
\end{equation}
A firm upper bound on the magnitude of
$\Delta$ is 
\begin{eqnarray}
|\Delta| \le {Q_b \over 2 \sigma_b}
         \le 0.5 q.
\label{equation-absolute-difference}
\end{eqnarray}
Thus, no difference is larger in magnitude than $q / 2$
standard deviations in the background. Provided $q$ is not
larger than $2$, we expect $\Delta$ to be roughly uniformly
distributed about zero (on discrete points corresponding to
integer multiples of $b_{\rm scale}/\sigma_b$) and so we expect the
mean and standard deviation of $\Delta$ (the mean and RMS
normalized difference) to be approximately
\begin{eqnarray}
\mu_{\Delta} \approx 0
\end{eqnarray}
and
\begin{eqnarray}
\sigma_{\Delta} \approx {q \over \sqrt {12}}
                  \approx 0.29 q.
\label{equation-rms-difference}
\end{eqnarray}
The quantity $\sigma_{\Delta}$ corresponds to the
square-root of the ``digital distortion'' $\mathcal{D}$
discussed by Gaztañaga et al.\ (2001). They show that a
quantization of a distribution of standard deviation
$\sigma_b$ will result in a distribution with a standard
deviation $\sigma_{\hat b}$ given by
\[
{ \sigma_{\hat b} \over \sigma_b} \approx  \sqrt{1 + {q^2 \over12}} \approx 1 +
0.144q,
\]
which comes from adding the variances in quadrature and
assuming that the difference $\hat b - b$ is uncorrelated
with $b$ (which is a good approximation provided $q$ is
small).


\section{Performance}
\label{section-performance}

This section presents tests and comparisons of the
performance of the quantization compression method proposed
here, qcomp, in terms of the distribution of differences,
compression ratio, and speed of compression and
decompression. Hcomp and optimal 16-bit and 32-bit encoders
serve as points of reference.

\subsection{Trial Images}

\begin{figure*}
\begin{center}
\includegraphics[height=0.2\linewidth,angle=270]{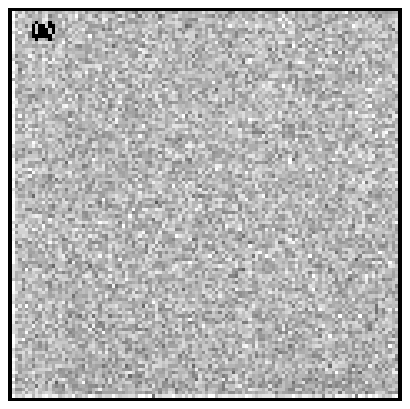}
~~~
\includegraphics[height=0.3\linewidth,angle=270]{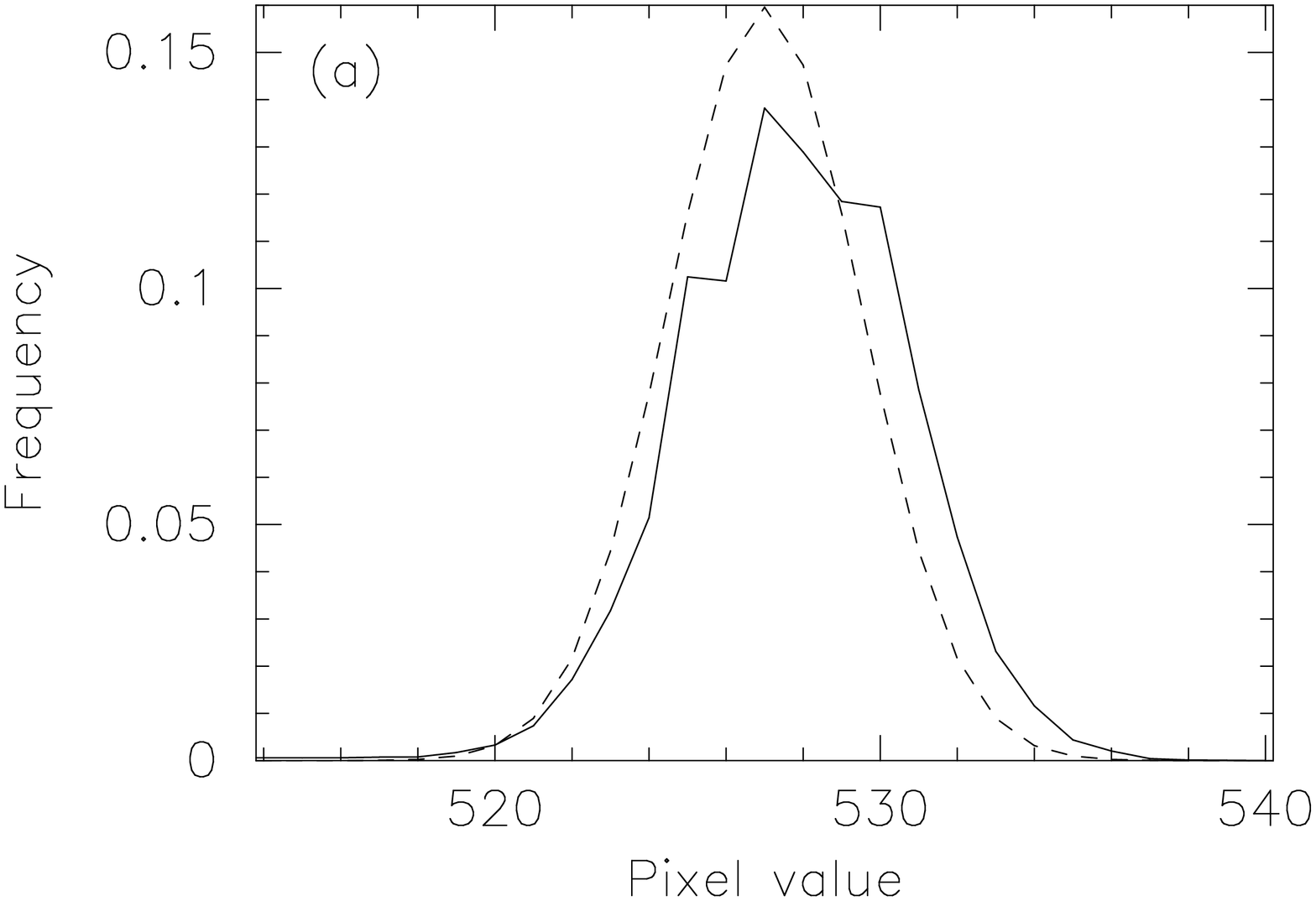}

\medskip

\includegraphics[height=0.2\linewidth,angle=270]{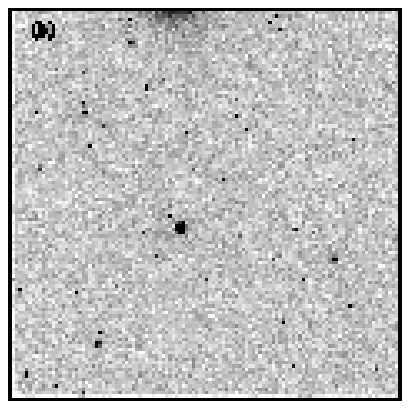}
~~~
\includegraphics[height=0.3\linewidth,angle=270]{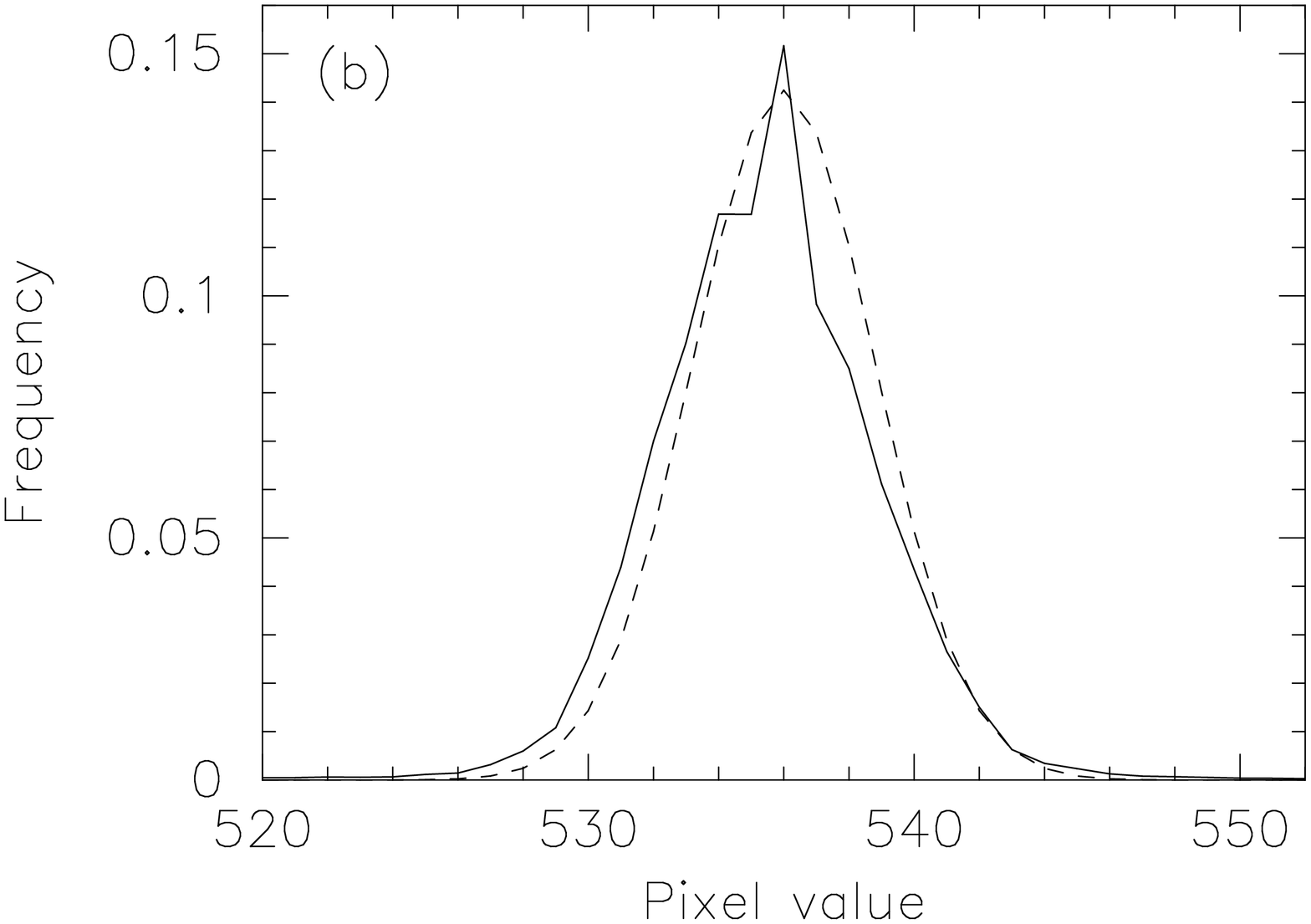}

\medskip

\includegraphics[height=0.2\linewidth,angle=270]{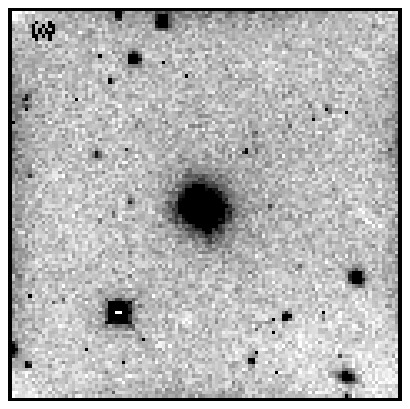}
~~~
\includegraphics[height=0.3\linewidth,angle=270]{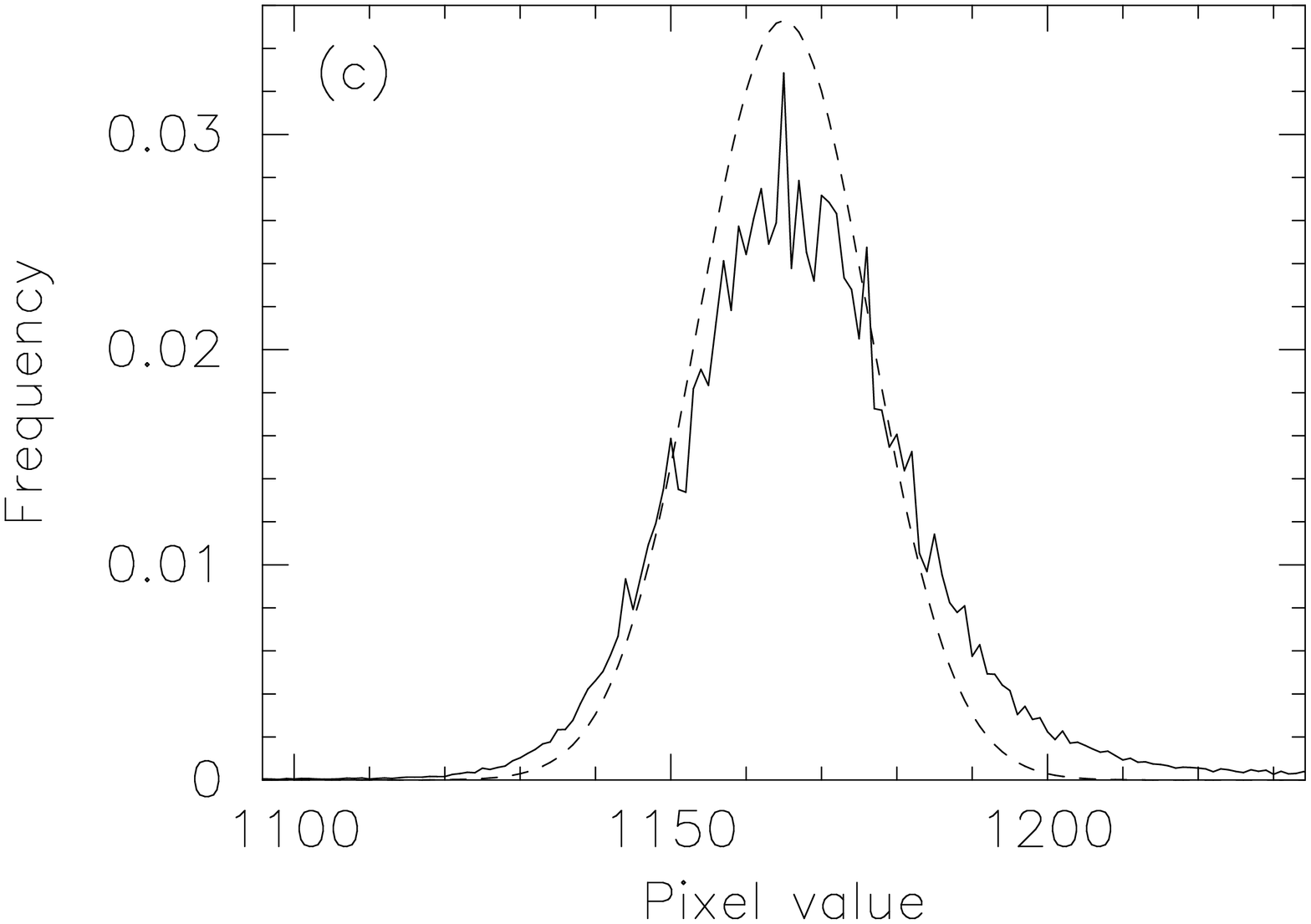}

\medskip

\includegraphics[height=0.2\linewidth,angle=270]{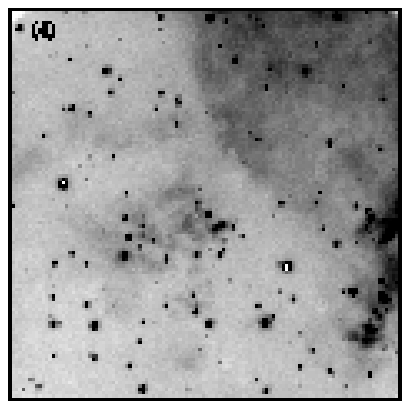}
~~~
\includegraphics[height=0.3\linewidth,angle=270]{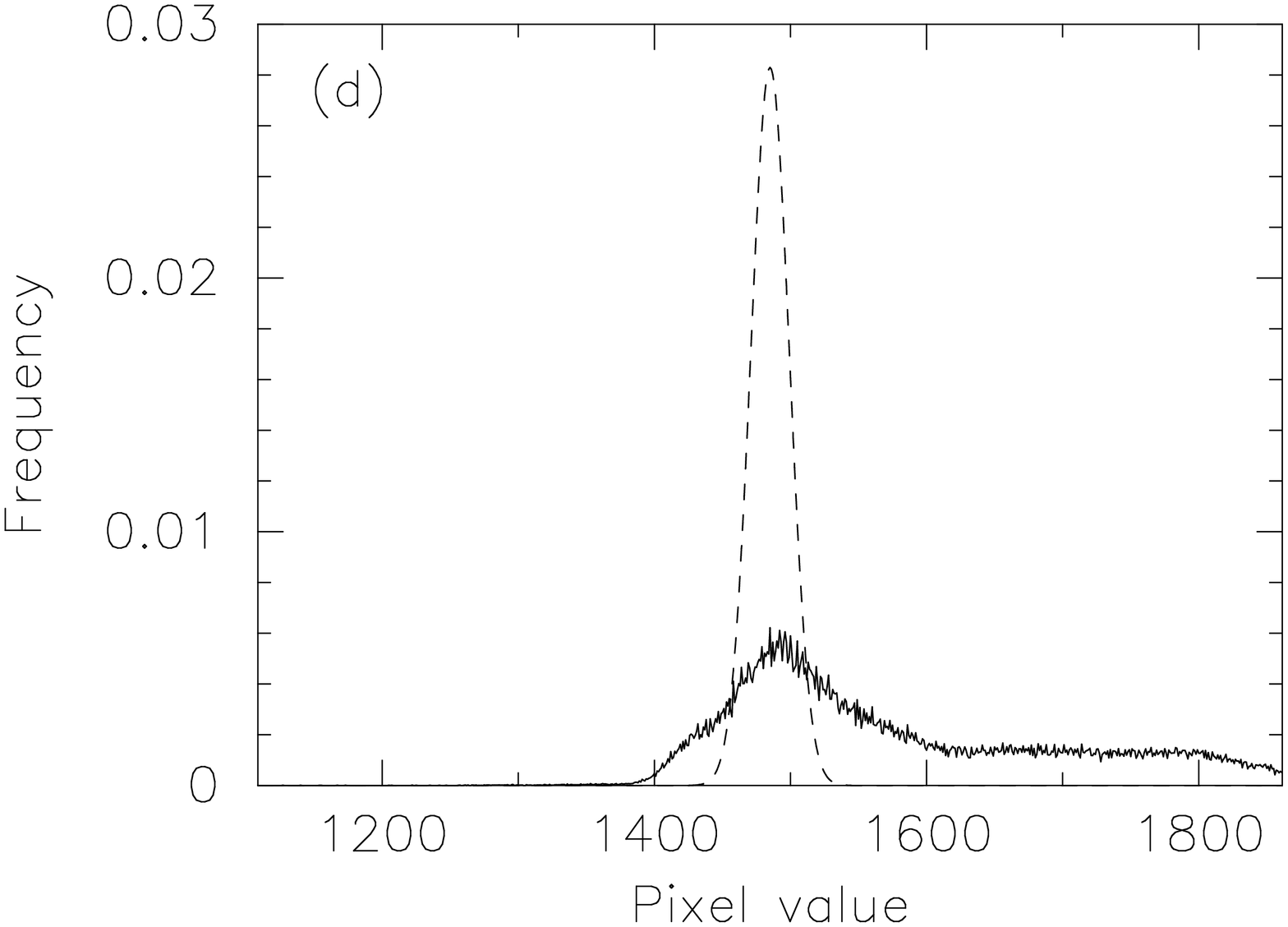}

\medskip

\includegraphics[height=0.2\linewidth,angle=270]{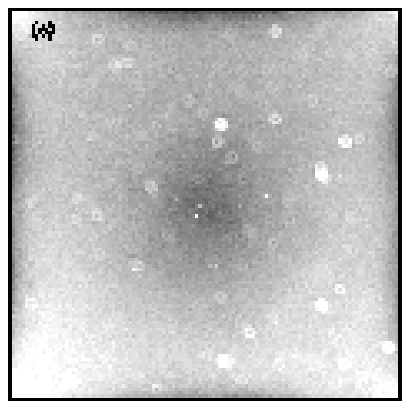}
~~~
\includegraphics[height=0.3\linewidth,angle=270]{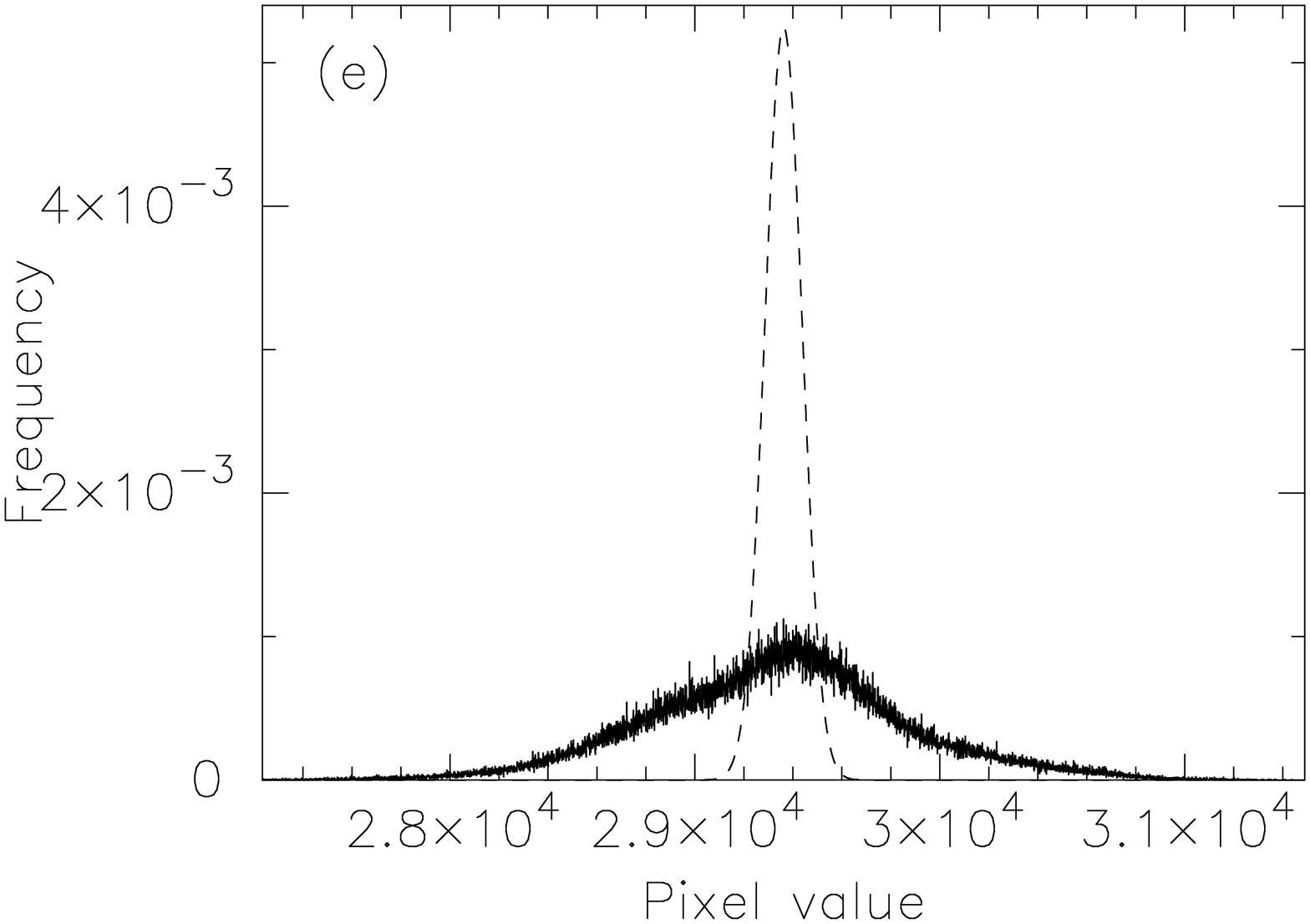}
\end{center}
\caption{The trial images. The images in the left panels are
stretched from $-2\hat\sigma_b$ to $+6\hat\sigma_b$ about
the mode and the histograms in the right panels are shown
stretched from $-4\hat\sigma_b$ to $+4\hat\sigma_b$ about
the mode, where $\hat\sigma_b$ is the standard deviation of
the pixel values (after 5$\sigma$ clipping). In the
histograms, the solid line is the observed distribution of
pixel values and the dashed line is the expected
contribution from noise. The real distributions are significantly
broader than the noise in image in image (d) because of
structure in the scene and in image (e) because of structure
in the flat field, both at low frequencies and on a
pixel-to-pixel basis.}

\label{figure-trial-images}
\end{figure*}

\begin{table}[tb]
\caption{Test Images}
\label{table-images}
\begin{center}
\begin{tabular}{llrr}
\hline\hline
Image   
&Description            
&\multicolumn{1}{c}{$\sigma_b$}
&\multicolumn{1}{c}{$\hat\sigma$}
\\
\hline
(a)     &Bias                   &2.5    &3.3\\
(b)     &Shallow sparse field   &2.8    &4.0\\
(c)     &Deep sparse field      &11.3   &17.4\\
(d)     &Deep complex field     &14.1   &94.1\\
(e)     &Flat field             &76.0   &532.0\\
\hline\hline
\end{tabular}
\end{center}
\end{table}

Five trial images were selected to sample different noise
regimes and scenes. The images were obtained with a CCD at
the Observatorio Astronómico Nacional at Sierra San Pedro
Mártir and saved as $512 \times 512 \times 16\mbox{-bit}$
FITS files with a BSCALE of 1. The images were (a) a bias
exposure, (b) a shallow exposure of a sparse field (a
standard star), (c) a deep exposure of a relatively sparse
field (a small galaxy), (d) a deep exposure of a complex
field (a nebula), and (e) a flat-field exposure. To make the
comparison with hcomp fair, the overscan sections were
removed, as hcomp does not distinguish between the data and
overscan sections.

The bias and background levels in each image were estimated
from the mode in the exposed and overscan regions, and the
noise $\sigma_b$ estimated by applying Equation
\ref{equation-noise-model}. The standard deviation
$\hat\sigma_b$ of the pixel values was also determined (with
$5\sigma$ clipping). Figure 2 shows the images and the
histograms of the pixel values, along with the expected
contribution contribution to the histogram width from the
noise. The histograms of the images (a), (b), and (c) are
dominated by noise. However, the histogram of image (d) is
much wider than the noise because of structure in the scene
and the histogram of image (e) is much wider because of
structure in the flat field (large-scale structure of around
1.5\% and pixel-to-pixel variations of around 0.5\%).

Any choice of trial images is likely to be somewhat
arbitrary. However, one advantage of these images is that
they cover a good range of noise regimes and scenes. This
implies that if a compression method can produce good
results for all of the trial images, it is likely to be
produce good results for other sets of images, regardless of
the relative frequency of each type of image. Thus, we see
that a successful compression method must achieve not only a
good mean compression ratio for the whole set of trial
images but also consistently good compression ratios for
each individual image.

\subsection{Compression}

The images were compressed using the following methods:

\begin{enumerate}

\item Lossless compression by hcomp with the ``-s 0''
option.

\item Lossless compression using hypothetical optimal 16-bit
and 32-bit methods. The compression ratios for these methods
were determined by calculating the Shannon entropy for each
image, given by Equation \ref{equation-shannon-entropy}.
Although these methods are hypothetical, they will
approximate Huffman and Arithmetic coding.

\item Lossless compression by direct application of bzip2,
gzip, and lzop with options varying from ``-1''
(fastest/worst) to ``-9'' (slowest/best).

\item Lossy compression using hcomp, with the degree of loss
being controlled by value of the ``-s'' option. The value of
this option was $2q\sigma_b$, which gives roughly the same
RMS difference as the quantization method with the same
value of $q$.

\item Lossy compression using the quantization method
described in \S~\ref{section-method} implemented by the
program qcomp (available from the author), with hypothetical
optimal 16-bit and 32-bit compression, bzip2, gzip, lzop,
and lossless hcomp used for the actual compression stage.

\end{enumerate}

\begin{figure*}[p]
\begin{center}
\includegraphics[height=\linewidth]{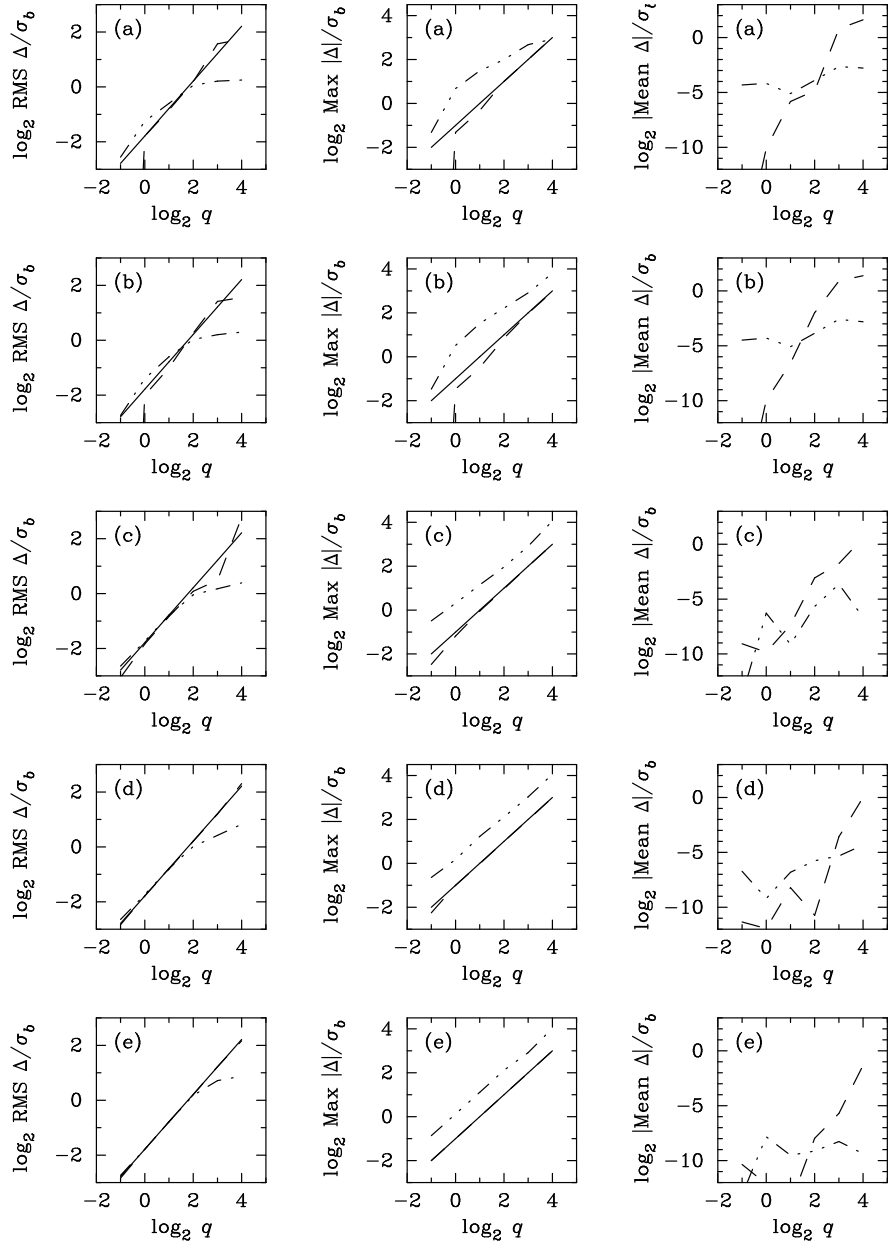}
\end{center}
\caption{The RMS, maximum absolute, and mean differences
between the compressed and original images as functions of
$q$ for the four images (a)--(e) for qcomp (dashed lines) and
hcomp (dotted-dashed lines). The solid lines show the relations
predicted for qcomp in \S~\ref{section-differences}.}
\label{figure-differences}
\end{figure*}

\begin{table*}[t]
\caption{Distributions of Differences}
\label{table-differences}
\begin{center}
\small
\begin{tabular}{llrrrrr}
\hline\hline
$q$     &
Method      &
\multicolumn{5}{c}{RMS, Maximum Absolute, and Mean
Differences $\Delta$ in units of $\sigma_b$}
\\
&
&\multicolumn{1}{c}{(a)}      
&\multicolumn{1}{c}{(b)}      
&\multicolumn{1}{c}{(c)}      
&\multicolumn{1}{c}{(d)}      
&\multicolumn{1}{c}{(e)}      
\\
\hline
0.5&qcomp    &0.00/0.00/$+$0.00 &0.00/0.00/$+$0.00 &0.12/0.18/$+$0.00 &0.14/0.21/$-$0.00 &0.14/0.25/$-$0.00\\
   &hcomp    &0.17/0.40/$+$0.05 &0.15/0.36/$+$0.04 &0.16/0.71/$-$0.00 &0.16/0.64/$+$0.01 &0.15/0.55/$+$0.00\\
\hline                                                                                                     
 1 &qcomp    &0.29/0.40/$-$0.00 &0.25/0.36/$-$0.00 &0.28/0.44/$+$0.00 &0.29/0.50/$-$0.00 &0.29/0.50/$-$0.00\\
   &hcomp    &0.42/1.60/$-$0.06 &0.38/1.43/$-$0.05 &0.30/1.24/$+$0.01 &0.30/1.13/$-$0.00 &0.29/1.09/$-$0.00\\
\hline                                                                                                     
 2 &qcomp    &0.56/0.80/$+$0.02 &0.50/0.71/$+$0.01 &0.56/0.97/$+$0.00 &0.57/0.99/$+$0.00 &0.58/1.00/$+$0.00\\
   &hcomp    &0.66/2.80/$+$0.03 &0.66/2.86/$-$0.03 &0.58/2.21/$+$0.00 &0.58/2.34/$-$0.01 &0.58/2.16/$-$0.00\\
\hline                                                                                                     
 4 &qcomp    &1.14/2.00/$-$0.03 &1.20/1.79/$-$0.25 &1.06/1.95/$-$0.12 &1.15/1.99/$+$0.00 &1.15/2.00/$+$0.00\\
   &hcomp    &1.04/4.00/$-$0.07 &1.02/4.64/$-$0.07 &0.97/3.89/$-$0.02 &1.01/4.33/$-$0.02 &1.13/4.33/$-$0.00\\
\hline                                                                                                     
 8 &qcomp    &2.97/4.00/$+$1.74 &2.67/3.93/$+$1.85 &1.40/3.98/$-$0.29 &2.29/3.97/$+$0.08 &2.34/4.00/$+$0.02\\
   &hcomp    &1.16/6.40/$-$0.16 &1.15/7.50/$-$0.17 &1.13/7.52/$-$0.08 &1.34/8.16/$-$0.02 &1.64/7.57/$-$0.00\\
\hline                                                                                                     
16 &qcomp    &3.34/8.00/$+$3.06 &2.95/7.86/$+$2.61 &6.85/7.96/$+$1.47 &4.95/7.94/$-$1.01 &4.48/8.00/$+$0.41\\
   &hcomp    &1.19/7.60/$-$0.15 &1.23/13.9/$-$0.14 &1.31/16.7/$-$0.01 &1.77/16.7/$-$0.05 &1.84/16.3/$-$0.02\\
\hline\hline
\end{tabular}
\end{center}
\end{table*}

\begin{table*}[p]
\renewcommand{\baselinestretch}{0.9}

\caption{Compression Ratios and Speeds}
\label{table-ratios-and-speeds}
\begin{center}
\footnotesize
\begin{tabular}{llcccccccc}
\hline
\hline
$q$     &
Method  &
\multicolumn{6}{c}{Compression Ratio}&
Compression   &
Decompression   
\\
&
&
(a)     &
(b)     &
(c)     &
(d)     &
(e)     &
mean    &
Mpixel/s        &
Mpixel/s        
\\
\hline
\multicolumn{10}{c}{\emph{Lossless methods}}\\
\hline
0   & hcomp            & 0.253 & 0.266 & 0.391 & 0.440 & 0.591 & 0.388 & \phantom{0}2.06 & \phantom{0}2.04 \\
    & optimal-16       & 0.232 & 0.246 & 0.395 & 0.561 & 0.692 & 0.425 \\
    & optimal-32       & 0.225 & 0.236 & 0.372 & 0.466 & 0.525 & 0.365 \\
    & bzip2 -1         & 0.242 & 0.256 & 0.393 & 0.460 & 0.634 & 0.397 & \phantom{0}0.73 & \phantom{0}2.07 \\
    & bzip2 -9         & 0.241 & 0.254 & 0.388 & 0.449 & 0.620 & 0.390 & \phantom{0}0.58 & \phantom{0}1.52 \\
    & gzip -1          & 0.329 & 0.343 & 0.502 & 0.641 & 0.769 & 0.517 & \phantom{0}2.81 & \phantom{}10.20 \\
    & gzip -9          & 0.296 & 0.313 & 0.488 & 0.621 & 0.766 & 0.497 & \phantom{0}0.36 & \phantom{}11.49 \\
    & lzop -1          & 0.498 & 0.509 & 0.709 & 0.828 & 0.992 & 0.707 & \phantom{0}9.80 & \phantom{}34.48 \\
    & lzop -3          & 0.498 & 0.509 & 0.709 & 0.825 & 0.991 & 0.706 & \phantom{}10.31 & \phantom{}34.48 \\
    & lzop -7          & 0.383 & 0.396 & 0.567 & 0.691 & 0.877 & 0.583 & \phantom{0}0.60 & \phantom{}32.26 \\
    & lzop -8          & 0.386 & 0.399 & 0.562 & 0.691 & 0.876 & 0.583 & \phantom{0}0.20 & \phantom{}32.26 \\
    & lzop -9          & 0.387 & 0.399 & 0.562 & 0.691 & 0.876 & 0.583 & \phantom{0}0.16 & \phantom{}32.26 \\
\hline
\multicolumn{10}{c}{\emph{Lossy methods}}\\
\hline
0.5 & hcomp             & 0.235 & 0.249 & 0.234 & 0.260 & 0.262 & 0.248 & \phantom{0}2.20 & \phantom{0}2.23 \\
    & qcomp/optimal-16  & 0.233 & 0.246 & 0.254 & 0.389 & 0.369 & 0.298 \\
    & qcomp/optimal-32  & 0.226 & 0.237 & 0.235 & 0.312 & 0.313 & 0.265 \\
    & qcomp/bzip2       & 0.243 & 0.257 & 0.246 & 0.274 & 0.280 & 0.260 & \phantom{0}0.83 & \phantom{0}2.46 \\
    & qcomp/gzip        & 0.330 & 0.343 & 0.341 & 0.421 & 0.448 & 0.377 & \phantom{0}3.52 & \phantom{}10.31 \\
    & qcomp/lzop        & 0.499 & 0.510 & 0.506 & 0.591 & 0.633 & 0.548 & \phantom{0}6.13 & \phantom{}32.26 \\
    & qcomp/hcomp       & 0.256 & 0.269 & 0.389 & 0.441 & 0.576 & 0.386 & \phantom{0}1.68 & \phantom{0}2.03 \\
\hline                                                                  
1   & hcomp             & 0.167 & 0.179 & 0.172 & 0.198 & 0.198 & 0.183 & \phantom{0}2.37 & \phantom{0}2.43 \\
    & qcomp/optimal-16  & 0.163 & 0.175 & 0.186 & 0.328 & 0.307 & 0.232 \\
    & qcomp/optimal-32  & 0.156 & 0.166 & 0.168 & 0.253 & 0.253 & 0.199 \\
    & qcomp/bzip2       & 0.176 & 0.189 & 0.176 & 0.209 & 0.214 & 0.193 & \phantom{0}0.85 & \phantom{0}2.62 \\
    & qcomp/gzip        & 0.259 & 0.271 & 0.268 & 0.352 & 0.373 & 0.305 & \phantom{0}3.83 & \phantom{}11.90 \\
    & qcomp/lzop        & 0.442 & 0.450 & 0.447 & 0.499 & 0.519 & 0.471 & \phantom{0}6.37 & \phantom{}32.26 \\
    & qcomp/hcomp       & 0.239 & 0.252 & 0.395 & 0.420 & 0.526 & 0.366 & \phantom{0}1.70 & \phantom{0}2.01 \\
\hline                                                                  
2   & hcomp             & 0.098 & 0.104 & 0.108 & 0.133 & 0.133 & 0.115 & \phantom{0}2.53 & \phantom{0}2.64 \\
    & qcomp/optimal-16  & 0.099 & 0.111 & 0.130 & 0.267 & 0.246 & 0.171 \\
    & qcomp/optimal-32  & 0.092 & 0.103 & 0.114 & 0.196 & 0.193 & 0.140 \\
    & qcomp/bzip2       & 0.108 & 0.120 & 0.120 & 0.148 & 0.154 & 0.130 & \phantom{0}0.83 & \phantom{0}2.76 \\
    & qcomp/gzip        & 0.186 & 0.197 & 0.204 & 0.279 & 0.292 & 0.232 & \phantom{0}4.31 & \phantom{}12.82 \\
    & qcomp/lzop        & 0.373 & 0.380 & 0.378 & 0.418 & 0.421 & 0.394 & \phantom{0}6.67 & \phantom{}35.71 \\
    & qcomp/hcomp       & 0.247 & 0.262 & 0.364 & 0.360 & 0.456 & 0.338 & \phantom{0}1.68 & \phantom{0}1.98 \\
\hline                                                                  
4   & hcomp             & 0.035 & 0.046 & 0.049 & 0.075 & 0.080 & 0.057 & \phantom{0}2.75 & \phantom{0}2.87 \\
    & qcomp/optimal-16  & 0.064 & 0.071 & 0.073 & 0.207 & 0.185 & 0.120 \\
    & qcomp/optimal-32  & 0.057 & 0.063 & 0.059 & 0.146 & 0.139 & 0.093 \\
    & qcomp/bzip2       & 0.070 & 0.075 & 0.063 & 0.097 & 0.101 & 0.081 & \phantom{0}0.53 & \phantom{0}3.37 \\
    & qcomp/gzip        & 0.137 & 0.144 & 0.122 & 0.203 & 0.185 & 0.158 & \phantom{0}5.13 & \phantom{}15.15 \\
    & qcomp/lzop        & 0.285 & 0.304 & 0.212 & 0.322 & 0.346 & 0.294 & \phantom{0}7.25 & \phantom{}37.04 \\
    & qcomp/hcomp       & 0.221 & 0.250 & 0.327 & 0.310 & 0.357 & 0.293 & \phantom{0}1.76 & \phantom{0}2.05 \\
\hline                                                                  
8   & hcomp             & 0.005 & 0.011 & 0.020 & 0.036 & 0.020 & 0.018 & \phantom{0}2.85 & \phantom{0}3.05 \\
    & qcomp/optimal-16  & 0.052 & 0.044 & 0.033 & 0.150 & 0.128 & 0.081 \\
    & qcomp/optimal-32  & 0.047 & 0.038 & 0.021 & 0.101 & 0.094 & 0.060 \\
    & qcomp/bzip2       & 0.060 & 0.049 & 0.015 & 0.060 & 0.060 & 0.049 & \phantom{0}0.31 & \phantom{0}3.73 \\
    & qcomp/gzip        & 0.120 & 0.099 & 0.030 & 0.127 & 0.122 & 0.100 & \phantom{0}5.68 & \phantom{}19.23 \\
    & qcomp/lzop        & 0.250 & 0.187 & 0.038 & 0.195 & 0.226 & 0.179 & \phantom{0}7.41 & \phantom{}38.46 \\
    & qcomp/hcomp       & 0.217 & 0.241 & 0.088 & 0.239 & 0.295 & 0.216 & \phantom{0}1.84 & \phantom{0}2.07 \\
\hline                                                                  
16  & hcomp             & 0.004 & 0.007 & 0.015 & 0.024 & 0.007 & 0.011 & \phantom{0}2.99 & \phantom{0}3.22 \\
    & qcomp/optimal-16  & 0.010 & 0.013 & 0.077 & 0.103 & 0.072 & 0.055 \\
    & qcomp/optimal-32  & 0.005 & 0.008 & 0.068 & 0.070 & 0.051 & 0.040 \\
    & qcomp/bzip2       & 0.002 & 0.004 & 0.072 & 0.041 & 0.030 & 0.030 & \phantom{0}0.40 & \phantom{0}3.51 \\
    & qcomp/gzip        & 0.007 & 0.011 & 0.140 & 0.083 & 0.061 & 0.060 & \phantom{0}5.52 & \phantom{}16.67 \\
    & qcomp/lzop        & 0.007 & 0.012 & 0.291 & 0.128 & 0.105 & 0.109 & \phantom{0}7.46 & \phantom{}38.46 \\
    & qcomp/hcomp       & 0.011 & 0.026 & 0.403 & 0.210 & 0.159 & 0.162 & \phantom{0}1.80 & \phantom{0}2.01 \\
\hline\hline
\end{tabular}
\end{center}
\end{table*}

\begin{figure*}[t]
\begin{center}
\includegraphics[height=\linewidth,angle=270]{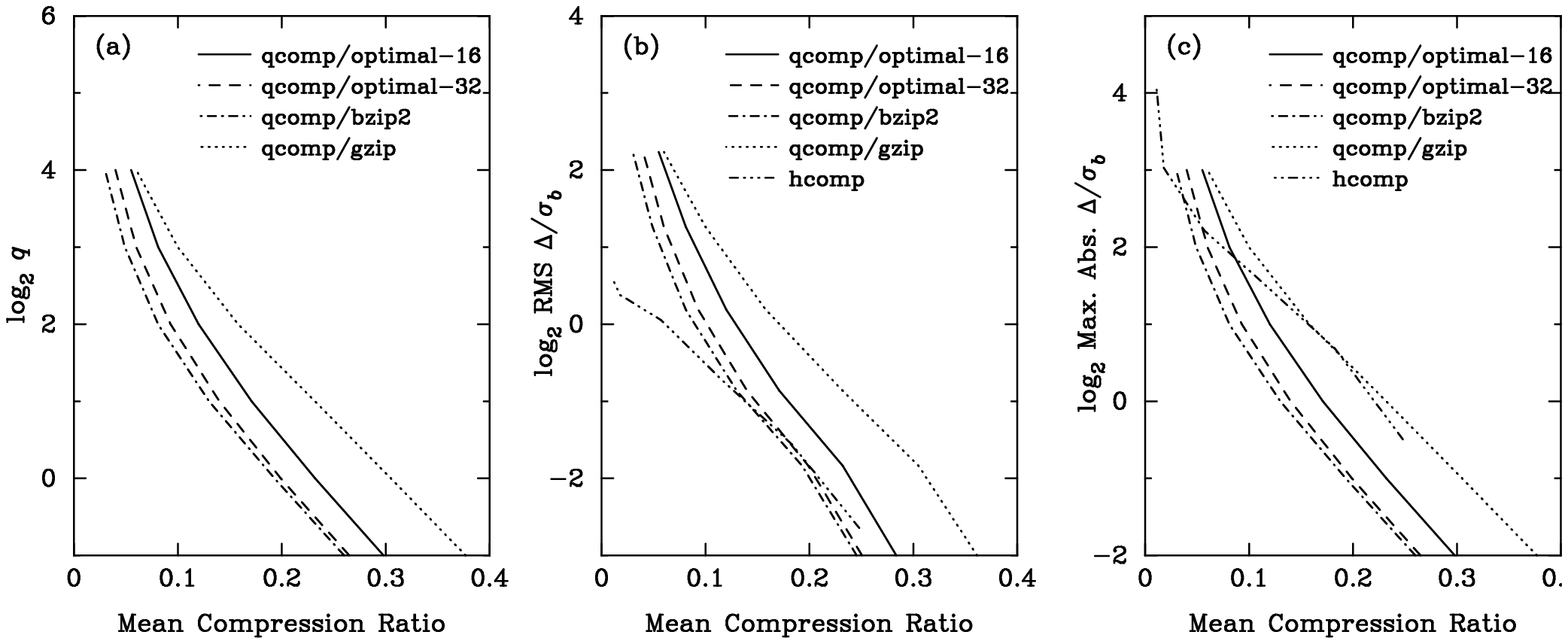}
\end{center}
\caption{The RMS and maximum absolute differences as
functions of the mean compression ratio for the lossy
compression methods. The solid line shows the Shannon
entropy limit for $q \le 1.5$.}
\label{figure-merit}
\end{figure*}

The distributions of differences, compression ratios, and
speeds are shown in Tables \ref{table-differences} and
\ref{table-ratios-and-speeds} and Figures
\ref{figure-differences} and \ref{figure-merit}. The lossy
methods are shown for $q$ values of 0.5, 1, 2, 4, 8, and 16.
Table \ref{table-differences} and Figure
\ref{figure-differences} show the RMS, maximum absolute, and
mean normalized difference $\Delta$ between the compressed
and original images. (The normalization is to $\sigma_b$,
the noise in the image; see Equation \ref{equation:delta}.)
Obviously, the lossless methods do not appear as their
differences are uniformly zero. Table
\ref{table-ratios-and-speeds} shows the compression ratio
achieved for each image (the ratio of the compressed size to
original size), the mean compression ratio for all five
images, and the speed at which the five images were
compressed and decompressed. These speeds were measured on a
computer with a single 1.9 GHz Pentium IV processor. Speeds
are not given for the optimal 16-bit and 32-bit methods, as
these methods are hypothetical.

The compression speeds quoted in Table
\ref{table-ratios-and-speeds} do not include the time taken
to determine the background level and calculate the noise.
Determining these using the mean, median, or mode proceeds
at about 11 Mpixel/s and reduces the compression speeds to
0.79 Mpixel/s for qcomp/bzip2 (9\% slower), 2.84 Mpixel/s
for qcomp/gzip (26\% slower), 4.03 Mpixel/s for qcomp/lzop
(37\% slower), 1.47 Mpixel/s for qcomp/hcomp (13\% slower),
and 1.94 Mpixel/s for hcomp (18\% slower) for $q = 1$.

Figure \ref{figure-merit} shows $q$, the RMS difference, and
maximum absolute differences as functions of the compression
ratio for the more effective lossy compression methods.

\subsection{Lossless Methods}

The upper section of Table \ref{table-ratios-and-speeds}
gives the compression ratios for the lossless methods. The
best methods are hcomp, bzip2, and the optimal 16-bit and
32-bit methods, which achieve mean compression ratios of
about 0.4.

In contrast to the compression of pure Gaussian noise
discussed in \S~\ref{section-pure-gaussian-noise}, hcomp and
bzip2 achieve better mean compression ratios than the
optimal 16-bit method. How can this be? The explanation is
that in the deep and flat-field images, (c), (d), and (e),
there are correlations between adjacent pixels that both
hcomp and bzip2 can exploit -- hcomp because its wavelets
can extend over more than one pixel and bzip2 because it can
increase the size of its input code unit to cover more than
one pixel. This also explains why the optimal 32-bit method,
which considers pixel in pairs, performs better than the
optimal 16-bit method, which considers pixels individually.
However, in the bias and shallow images, (a) and (b), the
noise component is dominant, and the optimal 16-bit method
performs slightly better than hcomp and bzip2, just as for
pure Gaussian noise. The optimal 32-bit method performs
slightly better than the optimal 16-bit method even for the
bias image, which suggests that there are small
pixel-to-pixel correlations even in this image, presumably
due to the bias structure. (Another way of looking at this
is that optimal 16-bit compression uses only the histogram
of pixel values whereas the other methods use both the
histogram and, to varying degrees, the image; the image adds
information on correlations between pixels, and allows for
more efficient compression.)

Nevertheless, the mean compression ratio of about 0.4 masks
variations of factors of 2--3 in the compression ratio for
individual images, with the bias and shallow images, (a) and
(b), being compressed well to about 0.25 and the deep and
flat-field images, (c), (d), and (e), being compressed
poorly to 0.4--0.7. This is as expected; the bias and
shallow images compress well as their noises are only
moderately over-sampled, but the deep and flat-field images
compress badly as their noises are significantly
over-sampled. In short, the lossless methods produce
inconsistent results, and while they will compress some sets
of data very well (for example, those dominated by
observations of standard stars or images in low-background
narrow-band filters), they will compress others very poorly
(for example, those dominated by deep images with high
backgrounds). In all, these results confirm what we already
knew: if we want to achieve consistently good compression
ratios, we have no choice but to adopt lossy methods.

(Table \ref{table-ratios-and-speeds} also shows the effect
of the options to bzip, gzip, and lzop. For bzip2 and gzip
there is not much improvement in the compression ratio
between the ``-1'' and ``-9'' options, but the compression
speed drops noticeably. For lzop there is a significant
improvement, but the compression speed drops precipitously.
It seems clear that the ``-1'' option is preferable, and for
this reason it was used in the lossless compression stage by
qcomp.)

\subsection{Lossy Methods}

Table \ref{table-differences} and Figure
\ref{figure-differences} shows that the distributions of
differences for qcomp are behaving much as predicted in
\S~\ref{section-differences}. For $q \le 2$, the RMS
difference is roughly $0.3q\sigma_b$, the maximum absolute
difference is $0.5q\sigma_b$, and the mean difference is
roughly zero. For larger values of $q$, the mean difference
ceases to be close to zero, as might be expected when the
quantization is too coarse to adequately sample both sides
of a sharply-peaked distribution. This suggests that qcomp
should not be used beyond $q \approx 2$.

Comparison of the distributions of differences for qcomp and
hcomp in Figure \ref{figure-differences} reveals two trends.
First, for $q \le 2$, the RMS differences are very similar.
This is not surprising as the hcomp option was set to give
this result. However, qcomp has lower values of the maximum
absolute difference and mean difference. For larger values
of $q$, the RMS differences and mean differences for hcomp
are much better than those for qcomp.

Moving from the performance of the quantization step to the
performance of the compression step, we first investigate
what appears to be a worrying anomaly: when the bias image,
(a), is quantized with $q = 1$, Table
\ref{table-ratios-and-speeds} shows that optimal 16-bit
compression gives a compression ratio of 0.163, whereas
Equation \ref{equation-shannon-entropy-for-gaussian}
indicates that optimal 16-bit compression of quantized
Gaussian noise with $q = 1$ should give a compression ratio
of about 0.128. The bias image has very little structure and
so should to a good approximation be simply Gaussian noise,
so why are the compression ratios so different? Only a small
part of the difference can be explained by corrections to
Equation \ref{equation-shannon-entropy-for-gaussian} for
finite $q$; numeric experiments give a true compression
ratio for Gaussian noise with $q=1$ of 0.134. The
explanation for the remaining difference is that the noise
$\sigma_b$ in the bias image is 2.5 and is not an integer
multiple of $b_\mathrm{scale}$, and so the quantum $Q_b =
b_\mathrm{scale}\lfloor q \sigma_b / b_\mathrm{scale}
\rfloor$ is rounded down to 2. Thus, the proper comparison
to compressing the bias image with $q = 1$ is compressing
Gaussian noise with $q = 0.8$. By Equation
\ref{equation-shannon-entropy-for-gaussian}, this would have
a compression ratio of 0.148, and numerical experiments give
a value of 0.153, which is much closer to the 0.163 of the
bias image. The remaining difference is presumably due to
structure in the bias and non-Gaussian contributions to the
noise (for example, the slightly flat top to the histogram
in Figure \ref{figure-trial-images}a). Similarly, the
shallow image (b) should be properly compared with Gaussian
noise quantized with $q=0.7$, which has a compression ratio
of 0.162. The other images have large values of $\sigma_b$,
so the truncation of $Q_b$ does not effect them
significantly.

Figure \ref{figure-merit}a shows the mean compression ratios
achieved by qcomp/optimal-16, qcomp/optimal-32, qcomp/bzip2,
and qcomp/gzip as functions of $q$. (Neither qcomp/lzop nor
qcomp/hcomp are shown, as they are significantly worse.) The
best mean compression ratios are achieved by qcomp/bzip2;
again, bzip2 appears to be finding correlations between
pixels that are not exploited by either the optimal 16-bit
or 32-bit methods, especially in the cases of the deep image
of the nebula (d) and the flat-field image (e). For $q = 1$
and $q = 2$ the mean compression ratios are 0.19 and 0.13,
that is, factors of two and three better than lossless
compression with bzip2. Furthermore, Table
\ref{table-ratios-and-speeds} shows that qcomp/bzip2 is
achieving consistently good compression ratios which vary by
only 20\% from 0.176 to 0.214 for $q = 1$.

Appropriate figures of merit for comparing hcomp and
qcomp/bzip2, are the RMS and maximum absolute difference at
a given mean compression ratio, and these are shown in
Figure \ref{figure-merit}b and c. For compression ratios
between 0.13 and 0.25 (which correspond to $q$ between 0.5
and 2), qcomp/bzip2 is better than hcomp both in RMS
difference and in maximum absolute difference. For
compression ratios smaller than 0.1 (which correspond to $q
> 2$), qcomp is significantly worse than hcomp, first in the
RMS difference but eventually also in the maximum absolute
difference. This is not surprising, as we saw above that the
quantization becomes too coarse for $q > 2$. Thus, in terms
of the distribution of differences at a given compression
ratio, of the methods we have considered, qcomp/bzip2 is
best for compression ratios of 0.13 and above (which
corresponds to $q \le 2$) and hcomp is best for compression
ratios smaller than 0.1 (which correspond to $q \ge 4$). The
one advantage that hcomp has over qcomp/bzip2 for $q \le 2$
is that it compresses about three times faster, although
both decompress at about the same speed.

At the same value of $q$ (i.e., the same distribution of
differences), qcomp/gzip2 produces significantly worse
compression ratios then qcomp/bzip2. For example, at $q = 1$
the mean compression ratio of qcomp/gzip is 0.31 compared to
0.19 for qcomp/bzip2. This appears to be not much better
than the 0.41 achieved by lossless compression by bzip2, but
qcomp/gzip has two advantages: it is roughly four times
faster then either lossless bzip2 or lossy qcomp/bzip2, both
at compression and decompression, and it achieves more
consistent compression ratios than lossless compression.

\section{Comparison to Other Quantization Methods}
\label{section-other-quantization-methods}

This section compares the quantization method proposed here,
qcomp, to the similar quantization methods proposed by White
and Greenfield (1999) and Nieto-Santisteban et al.\ (1999).

\subsection{Distribution of Differences}

White and Greenfield (1999) calculate the quantum from the
empirical standard deviation in the image rather than using
a noise model. This can cause the actual noise in raw data
to be significantly over-estimated. Consider the flat field
image (e), which has standard deviation of more than 500
(after 3 sigma clipping), but a noise of only about 75. Even
if we eliminate the contribution from large-scale structure,
the pixel-to-pixel standard deviation is about 160.
Obviously, as the empirical standard deviations is larger
than the noise, quantizing on the basis of the empirical
standard deviation would remove real information on the
variations in the flat field. Similar problems can occur in
any images with high backgrounds or more generally, in which
the empirical standard deviation is not dominated by noise.
This comparison is somewhat unfair, in that White and
Greenfield (1999) do not claim that their method is suitable
for raw data, but it does illustrate the danger of assuming
that the noise in an image is related to the empirical
standard deviation.

\begin{figure}[tb]
\begin{center}
\includegraphics[height=\linewidth,angle=270]{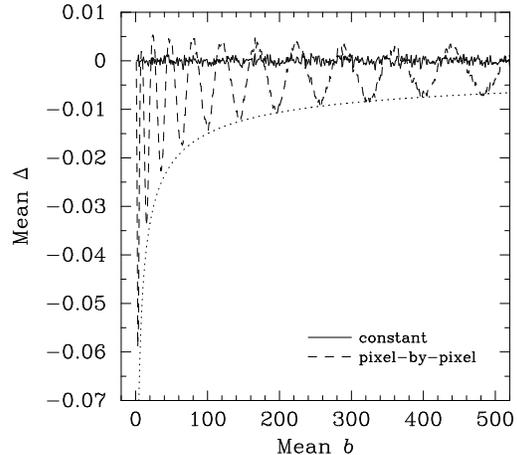}
\end{center}
\caption{The mean difference $\Delta$ (in units of the noise
$\sigma_b = \sqrt b$) between the compressed and quantized
images for images quantized with a single quantum for each
image and with quanta determined on a pixel-by-pixel basis.
The images have a gain of 1 and no read noise, and the
quantization has $q = 1$. The dotted line is the approximate
lower envelope to the pixel-by-pixel mean difference and has
the form $-0.15/\sigma_b$.}
\label{figure-pixel-by-pixel}
\end{figure}

Nieto-Santisteban et al.\ (1999) quantize by taking the
square root of scaled data. This is equivalent to
calculating the quantum on a pixel-by-pixel basis and
assuming that the read noise is zero. To investigate this, a
set of images were created with mean values from 1 to 200
electrons, a gain of 1 electron, no read noise, and with
appropriate Gaussian noise (although strictly speaking the
noise should be Poissonian), and written with BZERO of 0 and
BSCALE of 1. The images were quantized two different ways:
first using a single quantum for each image calculated by
applying Equations \ref{equation-noise-model} and
\ref{equation-x-quantization} with $q=1$ to the median pixel
value (the quantization method proposed here) and second
using individual quanta for each pixel calculated by
applying Equations \ref{equation-noise-model} and
\ref{equation-x-quantization} to each pixel value
(equivalent to the quantization method suggested by
Nieto-Santisteban et al.). The mean differences between the
quantized and compressed images are shown in Figure
\ref{figure-pixel-by-pixel}. Quantizing on a pixel-to-pixel
basis results in much larger mean differences than using a
single quantum, and the differences have an oscillatory
structure with a change of gradient every whole integer
squared. This probably results from the quantum changing
with pixel value, so that the two sides of the distribution
of pixel values are quantized slightly differently. The
lower envelope appears to be roughly $-0.15/\sigma_b$. The
mean difference is small even in the pixel-to-pixel case (of
order 1\% of the noise for typical signals), but
nevertheless quantizing with the same quantum gives a much
smaller mean difference (of order 0.05\% of the noise).

Thus, it appears that the quantization mehtod proposed here
has advantages over the other quantization methods when we
consider the fidelity of the compressed image to the
original image.

\subsection{Compression Ratio and Speed}

Rice, Yeh, \& Miller (1993) describe a special-purpose
compression algorithm designed to compress low-entropy
16-bit integer data. This method is widely referred to as
``Rice compression'' and has been used by White and
Greenfield (1999), Nieto-Santisteban et al.\ (1999), and, in
a modified form, White \& Becker (1998).

The compression ratio achieved by Rice compression appears
to be only very slightly better than that achieved by gzip;
White \& Becker (1998) report 0.255 for Rice and 0.272 for
gzip while Nieto-Santisteban et al.\ (1999) report 0.193 for
Rice and 0.202 for gzip (for 2 bits of noise).

White \& Becker (1998) report that gzip is ``far slower''
than their implementation of Rice compression and
Nieto-Santisteban et al.\ (1999) report that gzip is 20
times slower than White's implementation of Rice
compression. Both of these articles focus on compressing
data on a satellite with only limited computing resources,
and so understandably do not consider the speed of
decompression. Neither mention the flag given to gzip, but
the default is equivalent to ``-6'' and is about half as
fast at compression as ``-1''. If we assume that the default
was used, then Rice compression appears to be about 10 times
faster than gzip with the ``-1'' flag.

It would appear then that the methods proposed by White and
Greenfield (1999) and Nieto-Santisteban et al.\ (1999) would
produce similar compression ratios to qcomp/gzip2 but would
would compress much more quickly; neither method is likely
to produce compression ratios that are as good as those
produced by the much slower qcomp/bzip2. Of course, it would
be possibly to adapt qcomp to use Rice compression or the
other methods to use bzip2, in which case they should
produce very similar compression ratios and speeds.

\section{Discussion}
\label{section-discussion}

\subsection{Suitability for Compressing Raw Data}

The quantization compression method described here, qcomp,
was designed to compress raw data. With the results of the
tests and comparisons in \S~\ref{section-performance} and
\S~7, we can now critically evaluate its suitability for
this task. We will see that that qcomp is better suited for
compressing raw data than either hcomp (White 1992) or the
quantization methods proposed by White and Greenfield (1999)
and Nieto-Santisteban et al.\ (1999).

The fundamental advantage of qcomp over hcomp is that its
straightforward nature allows firm limits to be placed on
the distribution of differences (see
\S~\ref{section-differences}). So, for example, an image can
be compressed with the knowledge that no pixel will change
by more than a specified amount. Furthermore, the
distribution of the errors is predictable. Hcomp does give
such firm guarantees on the distribution of differences.

However, this advantage would be moot if qcomp were inferior
in compression ratio and speed. The most interesting range
of $q$ for compressing raw images is 0.5--2, that is,
quantizations that change the values in the image by at most
0.25--1 standard deviations. In this range, qcomp/bzip2
gives mean compression ratios of 0.27--0.13 and gives
roughly the same RMS difference as hcomp for similar
compression ratios but with smaller maximum absolute
differences and mean differences (see Figure
\ref{figure-merit}). Furthermore, its decompression speed is
very similar to hcomp. Its only disadvantage is that its
compression speed is only one third that of hcomp, although
it is still adequately fast; its compression rate of roughly
800 kpixel/s on a 1.9 GHz Pentium IV is much faster than the
typical read rate of 40 kpixel/s for single-output CCDs and
160 kpixel/s for quad-output CCDs.

It is also worth considering qcomp/gzip, which achieves
lower compression ratios of 0.38--0.23 for $q$ in 0.5--2,
while compressing slightly faster and decompressing very
much faster than hcomp. It also has the great advantage of
portability; gzip is installed almost universally on
workstations.

The current implementation of qcomp has a further small
advantage over the current implementation of hcomp in it
distinguishes between the data and overscan sections of an
image and quantizes each appropriately; the current
implementation of hcomp treats both identically, and so the
two have to be separated and compressed individually to
avoid either drastic over-compression of the overscan
section or under-compression of the data section.

The advantages of the qcomp over the other quantization
methods are that the manner in which White and Greenfield
(1999) determine the quantization makes their method
unsuitable for use with raw data (and indeed such a claim
was never made) and the manner in which Nieto-Santisteban et
al.\ (1999) quantize introduces a small bias that is not
present with qcomp. When using the same lossless compression
method, all of the quantization methods should be similar in
speed. The one disadvantage of qcomp is that is requires
that the noise in the background be roughly constant within
fixed regions, whereas the method of Nieto-Santisteban et
al.\ (1999) estimates the noise on a pixel-by-pixel basis
and makes no such requirement. However, this may not be too
restrictive as the noise varies only as the square root of
the background (and more slowly if read noise is
significant), so changes in the background produce much
smaller changes in the noise.

Obviously, confidence in the use of qcomp to compress raw
data would be significantly improved if end-to-end tests on
compressed data gave results that were statistically
indistinguishable from those derived from the original data.
Such tests should include astrometry of stars, aperture and
PSF-fitting photometry of stars, and photometry of low
surface brightness sources. These tests are especially
important for applications which have stringent requirements
on the absence of biases, such as far infrared and
sub-millimeter imaging and ultra low surface brightness
studies at other wavelengths. Such tests will be presented
in Paper II (Watson, in preparation).

\subsection{Distribution}

An important advantage of compressing files with qcomp/gzip
is that they can be distributed with the expectation that no
additional software will be needed to decompress them; gzip
is installed almost universally on workstations. As bzip2
becomes more widespread (it already forms part of several
Linux distributions), it too will gain this advantage. This
is a major advantage over software that uses special-purpose
compression methods.

\subsection{Media Capacities}

If we take a compression ratio of 0.2 as typical for
qcomp/bzip2 with $q=1$, then media have effective capacities
that are 5 times larger than their raw capacities. Thus, a
650 MB CD-ROM has an effective capacity of 3 GB, a 12 GB DAT
DDS-3 tape has an effective capacity of 60 GB, and an 80 GB
disk has a capacity of 400 GB. This is roughly equivalent to
a generation of technology; in terms of capacity, a CD-ROM
is almost a DVD-ROM, a DAT tape is effectively a DLT tape,
and a single disk is effectively a large RAID array. Thus,
compression can allow one to work with large data sets
without acquiring the devices often considered essential.

\subsection{Bandwidth}
\label{section-bandwidth}

\begin{table*}[tp]
\caption{Device and Transport Mechanism Speed}
\begin{center}
\begin{tabular}{lrrrr}
\hline
Device                  &\multicolumn{4}{c}{Bandwidth (Mpixel/s)}\\
                        &Raw    &Current Read  &Current Write  &Maximum    \\
\hline                                                                     
Slow single disk        &4      &12            &3              &13         \\
Fast single disk        &7      &12            &3              &23         \\
Fast RAID disk array    &25     &12            &3              &83         \\
$50 \times$ CD-ROM      &4      &12            &               &13         \\
$4 \times$ CD-R         &0.3    &              &1              &1          \\
DAT DDS-3               &0.6    &2             &2              &2          \\
DLT                     &2.5    &8             &3              &8          \\
56 kbps link            &0.003  &0.01          &0.01           &0.01       \\
10 Mbps link            &0.5    &1.7           &1.7            &1.7        \\
100 Mbps link           &5      &12            &3              &17         \\
1 Gbps link             &50     &12            &3              &150        \\
\hline
\end{tabular}
\end{center}
\end{table*}

If images can be compressed and decompressed sufficiently
quickly, compression increases the effective bandwidth of a
device or transport mechanism by a factor equal to the
inverse of the compression ratio. Thus, it may be faster to
store images in their compressed form, even though there is
an overhead in compressing or decompressing them.

Table 4 shows for several devices or transport mechanisms
the raw bandwidths and the current and maximum effective
bandwidths for compressed data, assuming a compression ratio
of 0.3 appropriate for compression with qcomp/gzip with
$q=1$. The maximum bandwidths assume an arbitrarily fast
processor, so the overheads of compression and decompression
drop to zero. The current read and write bandwidths were
measured or estimated for qcomp/gzip on an early-2002
computer with single 1.9 GHz Pentium IV processor, which can
compress at about 3 Mpixel/s and decompress at about 12
Mpixel/s. (A similar table could be constructed for
qcomp/bzip2; the maximum effective bandwidths would be
higher and the current effective bandwidths would be lower,
except for the 56 kbps link which would have a current effective
bandwidth of 0.015 Mpixel/s.)

Table 4 shows that even now compression results in an
improvement in effective bandwidth for CD-ROMs, tapes,
single disks, and all but the very fastest network
connections. As processors become faster, the actual
bandwidths will approach the maximum bandwidths, and, for
fast enough processors, all devices will benefit. For
example, achieving 83 Mpixel/s read bandwidth from a fast
RAID array will require processors only 7 times faster than
a 1.9 GHz Pentium IV; if processors continue to double in
speed every 18 months, according to Moore's law, such
processors should be available in about 2006. Achieving 83
Mpixel/s write speed will require processors 28 times faster
which should be available in about 2009. Alternatively, if
we allow 8-way parallelism (see \S~\ref{section-speed}),
such speeds should be possible now for reading and in about
2005 for writing. (Again assuming 8-way parallelism,
qcomp/bzip2 should give 125 Mpixel/s read bandwidth around
2006 and 125 Mpixel/s write bandwidth around 2008.)

\section{Possible Improvements}
\label{section-improvements}

\subsection{Compression Ratio}

The compression ratio could be improved by using a
compression method that performed better than bzip2.
However, the experiments above indicate that bzip2 is better
than both optimal 16-bit and optimal 32-bit compression, so
this may be difficult.

Furthermore, it is easy to show that bzip2 is performing
very close to optimally. The mean compression ratio achieved
by qcomp/bzip2 with $q=1$ is 0.193. The mean compression
ratios achieved by bzip2 and optimal 16-bit compression of
Gaussian noise with the same $q$ values (which, as mentioned
above, are really $q=0.8$ and $q=0.7$ for images (a) and
(b)) are 0.172 and 0.143. This suggests that a hypothetical
compression method that compressed the information as well
as bzip2 and compressed the noise as well as optimal 16-bit
compression would achieve a compression ratio of about
0.164, an improvement of only 18\%. Even a hypothetical
compression method that compressed the information to
nothing and compressed the noise as well as optimal 16-bit
compression would achieve a compression ratio of about
0.143, an improvement of only 35\%. It seems that bzip2 is
close to optimal, and any further improvements will be
small; there are no more factors of two to be gained.

One means to improve the compression ratio in
high-background images would be to roughly flatten and/or
subtract the sky prior to compression. This should narrow
the histogram of pixel values and thereby improve subsequent
compression. However, again, there's not much to be gained.
The difference between the compression ratios for the bias
image (a) and the flat-field image (e) is only 10\% for
qcomp/bzip2 with $q=1$; bzip2 is doing a very good job of
encoding both the pixel-to-pixel and large-scale structure
in the flat-field image, and so would presumably perform
similarly well on other high-background images.

\subsection{Speed}
\label{section-speed}

An obvious improvement to the writing speed would be to
parallelize the quantization and compression steps. This has
already been implemented in qcomp, as the quantized image is
piped to the lossless compressor as it is created. The
improvements are up to 50\% for qcomp/gzip and 10\% for
qcomp/bzip2.

To obtain further improvements, compression and
decompression must be made faster. For both gzip and
bunzip2, compression can be trivially parallelized by
splitting the input between processes and concatenating the
output from each process. Unfortunately, the file formats
for gzip and bzip2 do not permit decompression to be
parallelized. (The only way to determine the length of a
compressed block is decompress it.) However, it would be a
simple matter to modify the format of the compressed files
to permit decompression to be parallelized (by prefixing
each compressed block with its length); this would not be so
conveniently portable, but might be worthwhile for private
use.

Rice compression might be attractive if more speed is
required and compression ratios similar to those of gzip are
adequate. That said, even if the cost of lossless
compression could be reduced to zero, determining the noise
level and quantizing would still require effort and would
limit the compression speed to only about 3 times that of
qcomp/gzip. (It is possible that the noise level could be
determined more rapidly by determining the noise from only a
subsample of pixels, in which case the compression speed
might improve by a somewhat larger factor.)

\subsection{Robustness against Varying Backgrounds}

One problem with the current implementation of qcomp is that
it assumes that the whole data section of the image can be
adequately characterized by a single background and hence a
single background noise. (It does, however, allow the data
and bias sections to have different backgrounds.) This will
present a problem for images that have very different
background levels, for example, spectra in the thermal
infrared in which the thermal background increases rapidly
with wavelength.

This situation could be improved by subdividing the data
section into regions such that each region is adequately
characterized by a single background. This could be achieved
manually or perhaps adaptively. One adaptive scheme would be
to subdivide regions recursively until the backgrounds in
the subregions are not significantly different from the
background in the region.

Another option would be to estimate the noise and hence the
quantum on a pixel-by-pixel basis, following
Nieto-Santisteban et al.\ (1999), but as has been shown this
introduces a small bias in the mean difference between the
quantized and original images.

\section{Summary}
\label{section-summary}

A lossy method for compressing raw images has been
presented. The method is very simple; it consists of lossy
quantization (resampling in brightness) followed by lossless
compression. The degree of quantization is chosen to
eliminate the low-order bits that over-sample the noise,
contain no information, and are difficult or impossible to
compress.

The method is lossy but gives certain guarantees about the
distribution of differences between the compressed and
original images. In particular, it gives guarantees on the
maximum absolute value, the expected mean value, and the
expected RMS value of the difference. These guarantees make
it suitable for use on raw data.

The method consistently reduces images to 1/5 of their
original size while changing no value by more than 1/2 of a
standard deviation in the background. This is a dramatic
improvement on the compression ratios achieved by lossless
compression. The method is adequately fast on current
computers and would be relatively simple to parallelize.

A key feature of the method is that data can be uncompressed
using tools that are widely available on modern
workstations, which means that one can distribute compressed
data and expect that it can be used without the need to
install specialized software. This is achieved by writing
the quantized image as a normal FITS file and compressing it
with gzip and bzip2, which widely available general-purpose
compression tools. It appears that bzip2 is compressing the
data within a few tens of percent of optimally.

The next step in the development of this method is
real-world testing with compressed raw data to ensure that
the method does not degrade the results of astronomical
analyses. Such end-to-end tests will be presented in Paper
II (Watson, in preparation).

\acknowledgements

I thank Enrique Gaztañaga his referee's report, possibly the
most useful I have ever received, and in particular for his
suggestion to compare my results to the Shannon entropy
limit. I thank Rick White for useful comments on hcomp and
for blazing the trail. I thank Julian Seward, Jean-Loup
Gailly, Mark Adler, and Markus Oberhumer for doing the hard
work of writing bzip2, gzip, and lzop. I thank the staff of
the Observatorio Astron\'omico Nacional de M\'exico at San
Pedro M\'artir for their warm hospitality in February 1999,
when the ideas presented here were conceived.


\begin{thebibliography}

\bibitem{}Beckett, M.G., Mackay, C.D., McMahon, R.G.,
Parry, I.R., Ellis, R.S., Chan, S.J., \& Hoenig, M. 1998,
Proc.\ SPIE, 3354, 431

\bibitem{}Gailly, J.-L. 1993 ``gzip: The data compression
program'' (http://www.gzip.org/)

\bibitem{}Gaztañaga, E., Romeo, A., Barriga, J., \&
Elizalde, E. 2001, MNRAS, 320, 12

\bibitem{}Huffman, D.A. 1952, Proceedings of the Institute
of Radio Engineers, 40, 1098

\bibitem{}Louys, M., Starck, J.-L., Mei., S., Bonnarel, F.,
\& Murtagh, F. 1999, A\&AS, 136, 579

\bibitem{}Nieto-Santisteban, M.A., Fixsen, D.J., Offenberg,
J.D., Hanish, R.J. \& Stockman, H.S. in Astronomical Data
Analysis Software and Systems VIII, ed. D.M. Mehringer, R.L.
Plante, \& D.A. Roberts (San Francisco, ASP), p.~137

\bibitem{}Oberhumer, M.F.X.J. 1998, ``lzop - compress or
expand files''\\ (http://www.oberhumer.com/opensource/lzop/)

\bibitem{}Press, W.H. 1992, Astronomical Data Analysis
Software and Systems I, ed.\ D.M. Worrall, C. Biemesderfer,
\& J. Barnes (San Francisco, ASP), p.~3

\bibitem{}Press, W.H., Teukolsky, S.A., Vetterling, W.T., \&
Flannery, B.P. 1992, Numerical Recipes in C, (Cambridge:
CUP), 2nd edition

\bibitem{}Romeo, A., Gaztanaga, E., Barriga, J., \&
Elizalde, E. 1999, International Journal of Modern Physics
C, 10, 687

\bibitem{}Seward, J.R. 1998, ``bzip2 and libbzip2: a program
and library for data compression''
(http://sources.redhat.com/bzip2/)

\bibitem{}Shannon, C.E. 1948a, Bell System Technical
Journal, 27, 379

\bibitem{}Shannon, C.E. 1948b, Bell System Technical
Journal, 27, 623

\bibitem{}Shannon, C.E. 1949, Proceedings of the Institute
of Radio Engineers, 37, 10

\bibitem{}Veillet, C. 1998, ``MegaPrime: A New Prime-Focus
Environment and Wide Field Image Camera'' in CFHT
Information Bulletin, 39

\bibitem{}V\'eran, J.P., \& Wright, J.R. 1994, in
Astronomical Data Analysis Software and Systems III, ed.
D.R. Crabtree, R.J. Hanisch, \& J. Barnes (San
Francisco, ASP), p.~519
(ftp://www.cfht.hawaii.edu/pub/compfits/)

\bibitem{}Wells, D.C., Greisen, E.W., \& Harten, R.H.
1981, A\&AS, 44, 363

\bibitem{}White, R.L. 1992, in Proceedings of the NASA
Space and Earth Science Data Compression Workshop, ed.\ J.C.
Tilton (ftp://stsci.edu/software/hcompress/)

\bibitem{}White, R.L., \& Becker, I. 1998, ``On-board
compression for the HST Advance Camera for Surveys''
(http://sundog.stsci.edu/rick/acscompression.ps.gz)

\bibitem{}White, R.L., \& Greenfield, P. in Astronomical
Data Analysis Software and Systems VIII, ed. D.M. Mehringer,
R.L. Plante, \& D.A. Roberts (San Francisco, ASP), p.~125

\bibitem{}Whitten, I.H., Neal, R.M., \& Cleary, J.G. 1987,
Communications of the ACM, 30, 520

\end{thebibliography}
\end{document}